\documentclass[structabstract]{aa}  
\usepackage{amsmath}
\usepackage{natbib}
\usepackage{graphicx}
\usepackage{txfonts}
\bibpunct{(}{)}{;}{a}{}{,}  
\begin{document}
\title{Beryllium abundances in stars with planets:}
\subtitle{Extending the sample\thanks{Based on observations obtained
with UVES at VLT Kueyen 8.2 m telescope in programme 074.C-0134(A)}} 

\author{M.~C.~G\'alvez-Ortiz\inst{1} \and E.~Delgado-Mena\inst{2} \and
J.~I.~Gonz\'alez~Hern\'andez\inst{2,3} \and G.~Israelian\inst{2} \and
N.~C. Santos\inst{4} \and R.~Rebolo\inst{2,5} \and 
A.~Ecuvillon\inst{2}
}

\institute{Centre for Astrophysics Research, Science and Technology
Research Institute, University of Hertfordshire, Hatfield AL10 9AB,
UK. \email{M.Galvez-Ortiz@herts.ac.uk} 
\and
Instituto de Astrof\'isica de Canarias, 38200 La Laguna, tenerife,
Spain 
\and
Dpto. de Astrof\'{\i}sica y Ciencias de la Atm\'osfera, Facultad de
Ciencias F\'{\i}sicas, Universidad Complutense de Madrid, E-28040
Madrid, Spain   
\and
Centro de Astrof\'isica, Universidade do Porto, Rua das Estrelas,
4150-762 Porto, Portugal 
\and
Consejo Superior de Investigaciones Cient\'ificas, Spain} 

\date{Received November 2009; accepted }

\abstract
{Chemical abundances of light elements as beryllium in planet-host
stars allow us to study the planet formation scenarios and/or
investigate possible surface pollution processes.} 
{We present here an extension of previous beryllium abundance studies.
The complete sample consists of 70 stars hosting planets and
30 stars without known planetary companions. The aim of this paper is
to further assess the trends found in previous studies with less
number of objects. This will provide more information on the processes
of depletion and mixing of light elements in the interior of late type
stars, and will provide possible explanations for the abundance
differences between stars that host planets and ``single'' stars.}
{Using high resolution UVES spectra, we measure beryllium 
abundances of 26 stars that host planets and 1 ``single'' star mainly 
using the $\lambda$ 3131.065~{\AA}~\ion{Be}{ii} line, by fitting 
synthetic spectra to the observational data. 
We also compile beryllium abundance measurements of 44 stars hosting
planets and 29 ``single'' stars from the literature, resulting in a
final sample of 100 objects.} 
{We confirm that the beryllium content is roughly the same in stars
hosting planets and in ``single'' stars at temperatures $T_{\rm
eff} \gtrsim 5700$~K. The sample is still small for $T_{\rm eff} \lesssim 5500$~K, 
 but it seems that the scatter in Be abundances of dwarf stars is slightly 
 higher at these cooler temperatures.}
{We search for distinctive characteristics of planet hosts through
correlations of Be abundance versus Li abundance, age, metallicity 
and oxygen abundance. These could provide some insight in the
formation and evolution of planetary systems, but we did not find any
clear correlation.} 


\maketitle

\section{Introduction}

Since the discovery of the first extrasolar planets, many efforts have
been made to characterise planet-host stars and to find the features
that can distinguish them from stars that do not have any known
planetary companion \citep[e.g.][]{Santos04a,Fischer05}. 
The study of chemical composition and abundances in general
\citep[e.g.][]{Ecuvillon04a,Ecuvillon04b,Ecuvillon06,Gilli06,Neves09},
and, in particular, of the light elements Li, Be and B
\citep[e.g.][]{Santos02,Santos04c,Israelian04,Israelian09},
provided some hints on the influence of a planetary companion in
the composition and evolution of stars. These light elements are
destroyed by (p,$\alpha$)-reactions at relatively low temperatures
(about 2.5, 3.5 and $5.0~\times~10^6$~K for Li, Be and B,
respectively).
Thus, light elements and their abundance ratios are good
tracers of stellar internal structure and allow us to extract 
valuable information about the mixing mechanism and rotation 
behaviour in these stars \citep[e.g.][]{Pinsonneault90}.

Several groups \citep[e.g.][]{Gonzalez01,Santos01,Santos04a}  
have found a correlation between stellar metallicity and the
presence of giant planetary companion among solar-like stars. 
It has been also largely discussed whether the trend is due to
``primordial'' origin, i.e. the frequency of planetary companion is a
function of the proto-planetary disc metal composition 
\citep[see e.g.][]{Santos04a}; or is due to a posterior ``pollution'',
i.e. the metallicity excess is due to matter accretion after reaching
the main sequence \citep[see e.g.][]{Pasquini07}.   
Most studies suggest the "primordial" scenario although some
examples may indicate that both cases occur
\citep{Israelian01,Israelian03,Ashwell05,Laws03}. 

However, it is not clear whether a direct relationship between
metallicity and probability of planet formation exists or if it is
only applicable to the ``type'' of planets that have been  
discovered so far. \citet{Santos04a} suggested that two distinct
populations of exoplanets can be present depending on whether or not
the planet is formed by a metallicity-dependent process.
Recently, \citet{Haywood09} has also proposed that the metallicity
excess in the sample of stars hosting planets may come from a
dynamical effect of galactic nature (i.e. migration of stars in the
galactic disc), and may not be related with the giant planets
formation process. 

\citet{Israelian04} found that exoplanet hosts are significantly
more Li-depleted than comparison stars but only in the range
5600-5850 K. This result has been confirmed by \citet{Chen06},
\citet{Takeda07}, and \citet{Gonzalez08}. On the other hand,
\citet{Gonzalez08} proposed that planet-host stars hotter than
5900\,K have more Li than the comparison ones.  
There is observational evidence that Li depletion is
connected with the rotational history of the star
\citep[e.g.][]{Garcialopez94,Randich97}.
Thus, similar stars, i.e. with similar stellar parameters and age,
should have different depletion rates depending on the proto-planetary
disc mass and composition and its effect on the rotation of the parent
star \citep{Bouvier08}.
  
More recently, \citet{Israelian09} has confirmed that Li
is more depleted in Sun-like stars hosting planets than in similar
stars without detected planets. These stars with and without
planets, which fall in the effective temperature range 5700--5850~K,
do not show any differences in the correlation patterns of their Li
abundances with age and metallicity, pointing to a connection between
the presence of planets and the low Li abundance measured in Sun-like
planet-host stars. 

Whatever the metallicity "excess" in planet-host stars is due to, 
light elements can still provide insights on the influence of planets
on the parent star evolution and in the internal evolution of stars in
general, they serve as a key for understanding pollution ``events'',
and also as tracers of the possible planet-disc interaction. 

In previous works directly related with this paper
\citep{Santos02,Santos04b,Santos04c}, a comparison between stars
hosting planets and stars without known planets revealed that, with
some exceptions, the two samples follow approximately the same
behaviour in the variation of Be abundance with the temperature,
similar to Li trend. This behaviour shows a Be abundance maximum near
6100 K, decreasing towards higher and lower temperatures and a Be
"gap" for solar-temperature stars. 
\citet{Santos04c} also remarked the possible tendency of the planet
hosts to be more Be-rich, at least for $T_{\rm eff} \gtrsim
5700$~K.

In this paper we derive beryllium abundances for 26 stars with planets and one star
without planets, and collect the lithium abundances for most of
them. This sample enlarge the previous samples of 44 stars with
planets and 29 stars without known planetary companions reported in
\citet{Garcialopez98,Santos04c}. Thus, we end up with a sample of roughly 100 stars
where 70 host planets. We studied the behaviour of these abundances
with the spectral type and luminosity class in both samples of stars
in order to further assess the trends found before, and to study
possible interactions between extrasolar planets and their parent
stars and its connection to the light-element abundances through 
time. 

\begin{figure}[!ht]
\centering
\includegraphics[width=9.2cm,angle=0,clip]{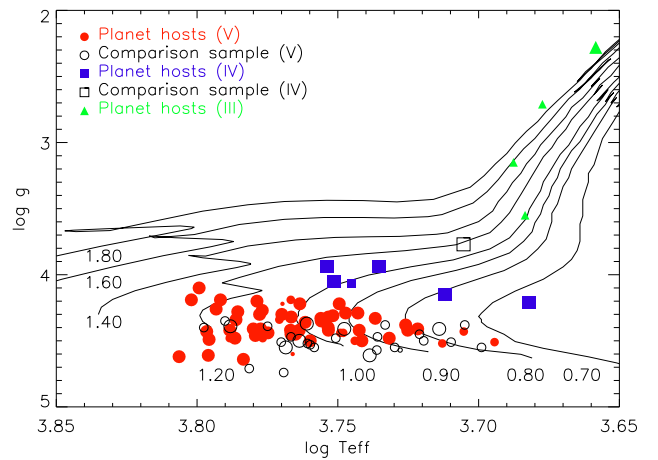}
\caption{HR diagram: surface gravity vs effective temperatures of the stars in
our sample including the stars in Tables~\ref{tab:par},~\ref{tab:parb}
and~\ref{tab:parc}. 
We overplot solar metallicity tracks from \citet{Girardi99} for 0.7 to
1.8 M$_{\odot}$. The luminosity classes of the stars in the sample are
assigned according with each star position in this diagram (See
Sect.~\ref{secsample}). Metallicity is represent by the size of 
the symbol.}
\label{fig:hr}
\end{figure}

\begin{figure*}[!ht]
\centering
\includegraphics[width=8.5cm,angle=0]{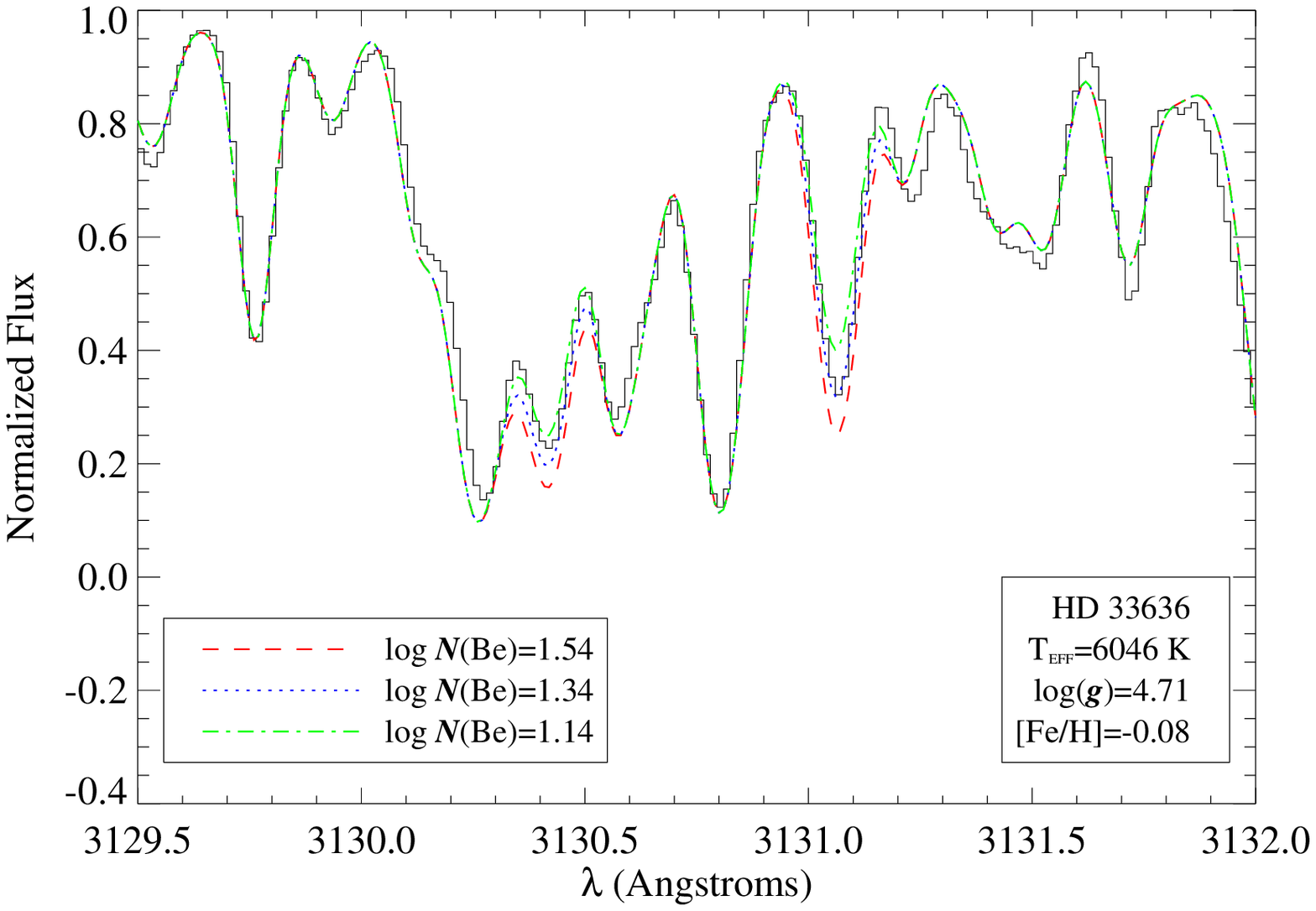}
\includegraphics[width=8.5cm,angle=0]{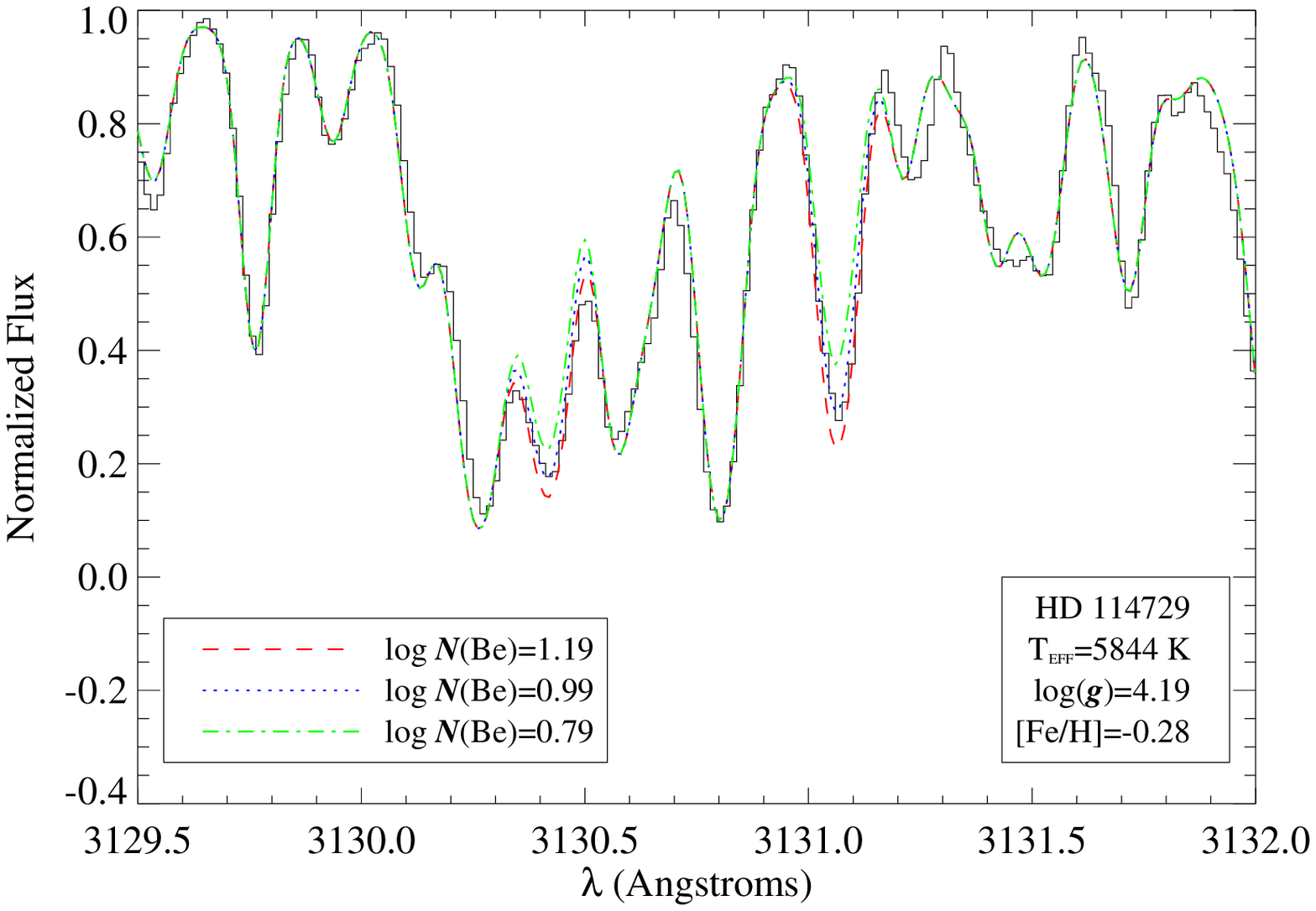}
\includegraphics[width=8.5cm,angle=0]{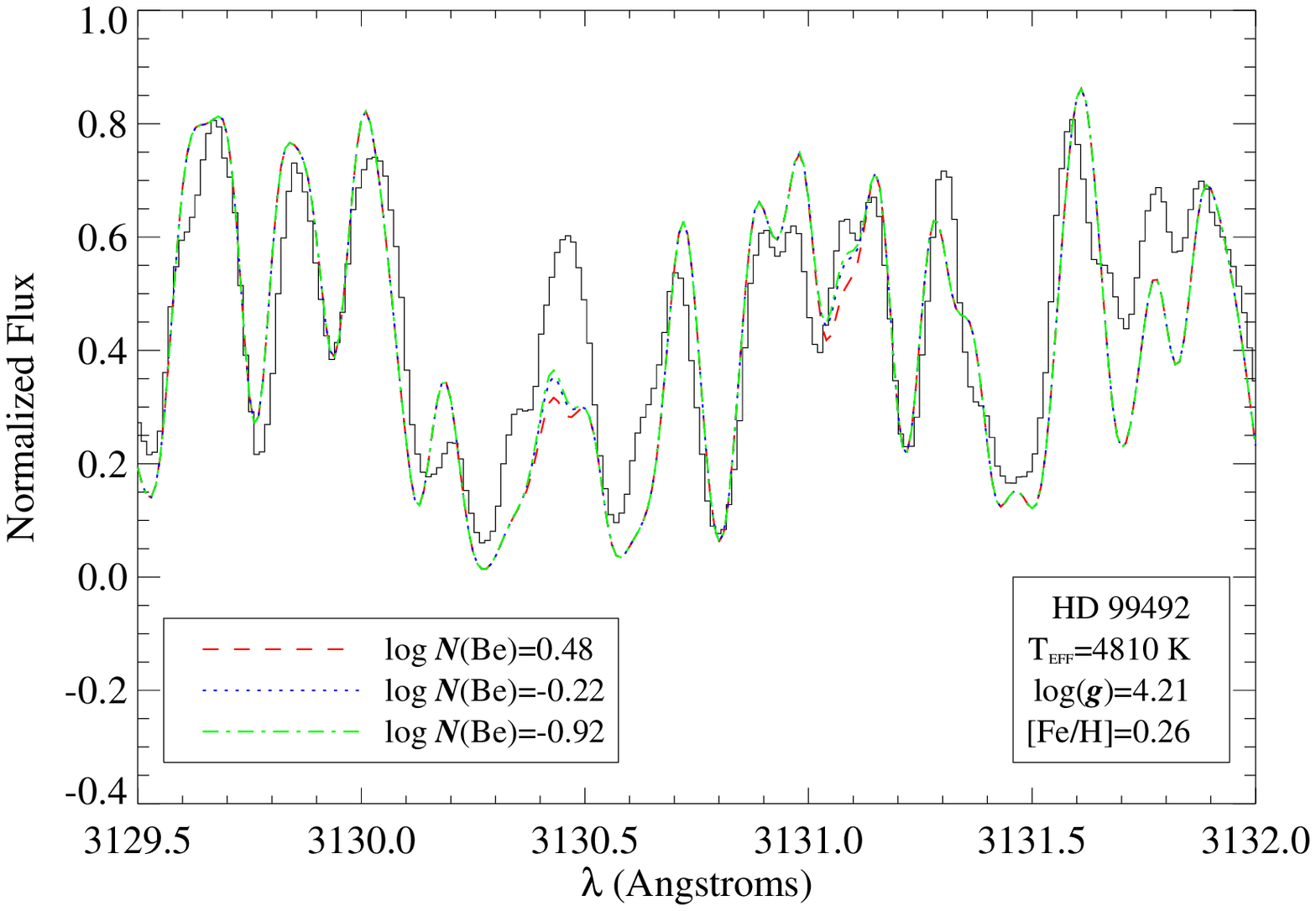}
\includegraphics[width=8.5cm,angle=0]{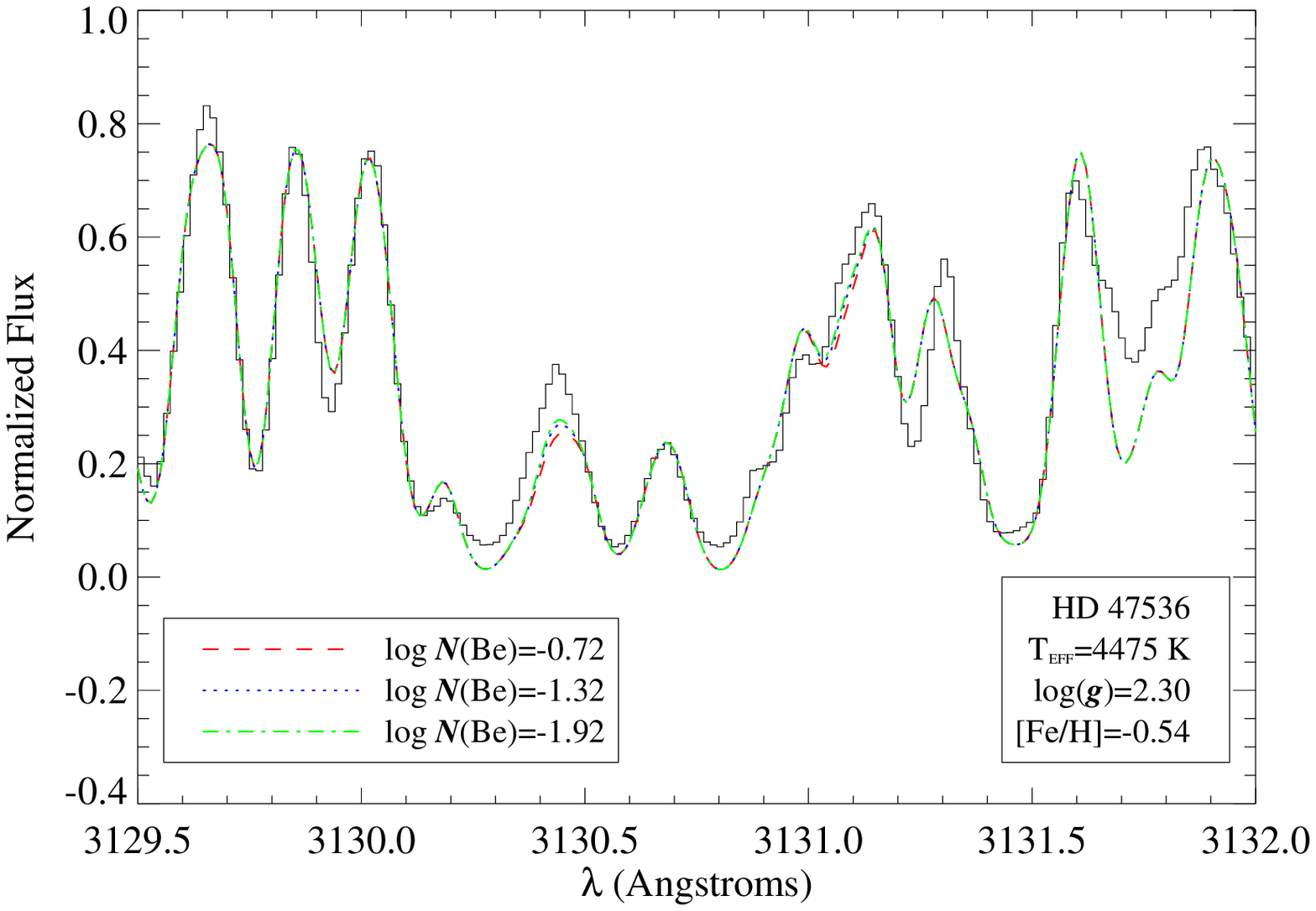}
\caption{Best synthetic spectral fits to the observed spectra in the \ion{Be}{ii}
line region. From upper to lower from left to right, the planet-host
dwarfs HD~33636 and HD~114729, the subgiant HD~99492 and the giant
HD~47536 are displayed.} 
\label{figfits}
\end{figure*}

\section{Observations and data reduction}

We obtained near-UV high-resolution spectra of the targets
using the UVES spectrograph at the 8.2-m Kueyen VLT (UT2)
telescope (run ID 074.C-0134(A)) on 21, 22 December 2004, 
in the blue arm with a wavelengh coverage of 
$\lambda\lambda$3025-3880~{\AA}. These spectra have a spectral
resolution $\lambda/\delta\lambda\sim70,000$. 

All the data were reduced using IRAF\footnote{IRAF is distributed by 
the National Optical Observatory, which is operated by the Association
of Universities for Research in Astronomy, Inc., under contract with
the National Science Foundation.}  tools in the echelle package.
Standard background correction, flat-field, and extraction procedures
were used. The wavelength calibration was done using a ThAr lamp
spectrum taken during the same night. Finally, we normalized the
spectra by a low-order polynomial fit to the observed continuum.

\section{Sample\label{secsample}}

The total sample of $\sim 100$ stars, including 70 planet-host stars,
cover a wide temperature range from 4500 to 6400~K. 
Thus, the stars hosting planets have spectral types between F6 to K2
and three luminosity classes (V, IV and III). The luminosity classes
were assigned following a criterion based on the position of the star
in the Hertzsprung-Russell (HR) diagram, and also on their surface
gravity, i.e. dwarfs (V) have typically $\log g > 4$, subgiants (IV)
have $3.3<\log g<4$ and giants (III) have $\log g < 3.3$. 
We built an effective temperature-surface gravity plot with solar 
metallicity tracks from \citet{Girardi99} for 0.7 to 1.8 M$_{\odot}$
and according to each star position in this diagram
(see Fig.~\ref{fig:hr}), we assigned a 
luminosity class, presented in column 10 of Table~\ref{tab:par}
and column 7 of Tables~\ref{tab:parb} and ~\ref{tab:parc}. 
In Fig.~\ref{fig:hr} we represent the metallicity of the objects by 
increasing the symbol size according to three possible metallicity
ranges from smaller to bigger size: $[{\rm Fe}/{\rm H}]<-0.5$,
$-0.5<[{\rm Fe}/{\rm H}]<0.0$ and $0.0<[{\rm Fe}/{\rm H}]<0.5$, 
respectively. Thus, our sample of planet-host stars  with new Be 
abundance measurements contains 3 giants, 2 subgiants and 21 
dwarfs that give a total of 4 giants, 6 subgiants and 
 60 dwarfs in this study. 
The 30 stars without known planets of our sample cover the same 
spectral types but only two of them are a subgiant star. 
All the new objects are provided in Table~\ref{tab:par} 
(with 26 stars with planets and one star without known planetary 
companions). Stars with and without known planets from literature 
are listed in Tables ~\ref{tab:parb} and ~\ref{tab:parc}, 
respectively.

\section{Stellar parameters\label{secpar}}

Stellar parameters were taken from the detailed analysis of
\citet{Sousa08} when available, otherwise from \citet{Santos04a, 
Santos05}. These parameters are determined
from ionization and excitation equilibrium of \ion{Fe}{i} and
\ion{Fe}{ii}. They used the 2002 version of the code MOOG\footnote{The
source code of MOOG 2002 can be downloaded at 
http://verdi.as.utexas.edu/moog.html} \citep{Sneden73} and a grid of
local thermodynamical equilibrium (LTE) model atmospheres
\citep{Kurucz93}. 
The adopted parameters, effective temperature, $T_{\rm eff}$, surface
gravity, $\log g$, metallicities, [{\rm Fe}/{\rm H}], 
 and masses are listed in Table~\ref{tab:par}. 
The mean values of the uncertainties on the parameters from 
\citet{Santos04a, Santos05} are of the order of 44 K for 
$T_{\rm eff}$, 0.11~dex for $\log g$, 0.08~km~s$^{-1}$ for $\xi_{t}$, 
0.06~dex for metallicity and the adopted typical relative error
for the masses is 0.05 M$_{\odot}$.  
Whereas from \citet{Sousa08} these mean uncertainties are 25 K for 
$T_{\rm eff}$, 0.04 dex for $\log g$, 0.03 km s$^{-1}$ for 
$\xi_{t}$ and 0.02 dex for metallicity and 0.10 M$_{\odot}$ for
the masses.  
We refer to \citet{Sousa08} and \citet{Santos04a, Santos05} for 
further details. 
In Tables~\ref{tab:parb} and ~\ref{tab:parc} we give the main
data of the sample of stars with and without planets from
\citet{Santos04b,Santos04c}. We note here the uniformity of the
adopted stellar parameters \citep[see Section 5
in][]{Sousa08}. 

 One should note that stellar masses are estimated from 
theoretical isochrones which are strongly sensitive to the
adopted helium and metal content of the star. 
In addition, the stellar mass, although being a 
fundamental parameter, is relatively more uncertain that other 
stellar parameters, because it depends on the effective 
temperature, surface gravity and metallicity of the star. 
However, unlike the stellar age,
the stellar mass for main-sequence stars is typically better 
constrained from theoretical isochrones than for subgiant and/or 
giant stars \citep[see e.g.][]{Allende04}.

\begin{figure}[!ht]

\centering
\includegraphics[width=6cm,angle=270,clip]{figure6_go.ps}
\includegraphics[width=6.0cm,angle=270,clip]{bemas12o.ps}
\caption{Be abundances as a function of effective
temperature (top panel) and as a function of mass 
(bottom panel) for the planet-host stars in our sample plus the 
stars from \citet{Garcialopez98,Santos02,Santos04b,Santos04c}. 
Filled and open symbols represent stars with and without planets. 
Circles, squares and triangles, are depicted for dwarfs (luminosity 
class V), subgiants (IV) and giants (III). In the top panel 
we superimpose the Be depletion isochrones (case A) of
\citet{Pinsonneault90} for solar metallicity and an age of 1.7 Gyr.
From top to bottom, the lines represent a standard model (solid line)
and 4 models (dashed lines) with different initial angular momentum.
We also depict the Be depletion model (dotted line) including gravity
waves provided in \citet{Montalban00}.}
\label{figbeteff}
\end{figure}

\begin{figure}[!ht]
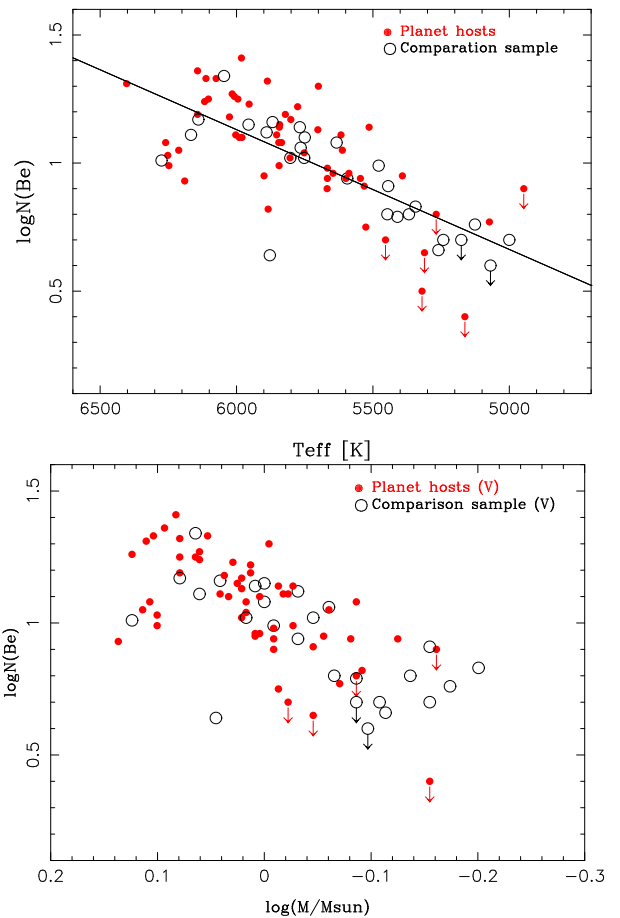

\centering
\includegraphics[width=6cm,angle=270,clip]{betefig4_29oct.ps}
\includegraphics[width=6cm,angle=270,clip]{bemasfig4_290ct.ps}
\caption{Be abundances as a function of effective temperature 
 (top panel) and as a function of mass (bottom panel) for 
only dwarfs with (filled circles) and without known planetary
companion (open circles). Overplotted are the best linear fits found
for both population together.}
\label{figbefit}
\end{figure}

\section{Beryllium and lithium abundances\label{secBeLi}}

The Be abundances were derived by fitting synthetic spectra to the
data. These synthetic spectra were convolved with a Gaussian smoothing 
profile and a radial-tangential profile to take into account 
the spectral resolution and the macroturbulence, respectively. 
The latter was varied between 1.0 and 5.0~km~s$^{-1}$, between K and F
dwarfs \citep{Gray92}, respectively. A rotational profile was also
added to account for the projected rotational velocity, $v\sin i$, of
the sample stars, given in Table~\ref{tab:par}. 
These were calculated from the width of the CORALIE
cross-correlation function \citep[see appendix of][]{Santos02}, or
taken from the literature \citep[mostly][]{Fischer05}. 
A limb-darkening coefficient of 0.6 was adopted in all cases,
and the overall metallicity was scaled to the iron abundance.

To derive Be abundances, we fitted all the observed spectral range 
between 3129.5 and 3132~{\AA}, where two \ion{Be}{ii} lines are
located ($\lambda$3130.420 and $\lambda$3131.065 {\AA}). 
We used mainly the \ion{Be}{ii} $\lambda$3131.065~{\AA} line as 
the feature for deriving the best-fit Be abundance, and 
\ion{Be}{ii} $\lambda$3130.420~{\AA} was used only for checking, 
since it is highly blended with other elements lines. 

We adopted the same line list as in \citet{Santos04c}, which
provides a solar photospheric Be abundance of $\log N({\rm
Be})=1.1$ ~dex\footnote{Here we use the notation
$\log N({\rm Be})=\log [N({\rm Be})/N({\rm H})]+12$}, using the same
models and tools as in this work. The main source of error
is probably the placement of the continuum. An uncertainty on the
continuum position of $3\%$ yields to an error in the Be abundance of
$\sim 0.05$~dex for solar type stars \citep[see e.g.][]{Randich02}.
We estimate this uncertainty to be $\sim 0.06$~dex at $T_{\rm
eff}=5500$~K, $\sim 0.11$~dex at $T_{\rm eff}=5200$~K, 
and $\sim 0.20$~dex at $T_{\rm eff}=4900$~K.
Another weak point is
the possible 0.3~dex difference between the solar photospheric Be
abundance and the meteoritic Be abundance. It has been suggested that
this could be due to our inability to properly account for all
continuum opacity sources in the UV \citep{Balachandran98}. However,
other authors have argued that Kurucz atmospheric models are able to
reproduce the near-UV absolute continuum, at least, for
stars in the $T_{\rm eff}$ range 4000--6000~K \citep{Allende00}.
In fact, \citet{Balachandran98} argue that the \ion{Fe}{i} bound-free
opacity should be increased by a factor of 1.6, which is equivalent to
an increase of 0.2~dex in the Fe abundance. \citet{Smiljanic09} have
tested this possibility for a model of [{\rm Fe}/{\rm H}]~$-0.5$~dex 
and they find that the difference in Be abundance is small, 
$\Delta\log N({\rm Be})\sim 0.02$~dex. 

For stars with effective temperatures below 5100~K, the spectral
region surrounding the \ion{Be}{ii} $\lambda$3131~{\AA} line begins to
be dominated by the contribution of other species (\ion{Mn}{i}, C, OH
lines, etc.) and for the giant stars also the thulium line 
(\ion{Tm}{ii} $\lambda$3131.255~{\AA}) may play an
important role \citep[see][]{Melo05}. This makes it more
difficult to fit the observed spectra and thus, to measure Be
abundance. Although we took into account some of these other  
elements when fitting cooler and giant objects, we assume that Be
abundances are not so accurate in these cases. In addition, 
the sensitivity of the \ion{Be}{ii} to the Be abundance
decreases towards lower temperatures. Thus, the total error,
including mainly the uncertainties on the effective temperature and
the continuum location, in the Be abundance is of the order of 
0.1~dex at $T_{\rm eff}\sim 6000$~K, 0.12~dex at 
$T_{\rm eff}\sim 5500$~K, 0.17~dex at 
$T_{\rm eff}\sim 5200$~K, and almost 0.3~dex at 
$T_{\rm eff}\sim 4900$~K. 

All the Be abundances measured in this work are listed in
Table~\ref{tab:par}. We have also added the Be abundance
measurements of the 44 stars hosting planets and 
29 ``single'' stars from previous studies 
\citep{Garcialopez98,Santos02,Santos04b,Santos04c}, listed 
in Tables \ref{tab:parb} and~\ref{tab:parc}.
We have compiled Li abundance measurements, obtained from the
\ion{Li}{i} $\lambda$6708~{\AA} line, from the literature
\citep{Israelian04,Israelian09}. We also measure new Li 
abundances in four objects (three giants and a subgiant, see 
Table~\ref{tab:par}) in the same way as in 
\citep{Israelian04,Israelian09}. 
In Fig.~\ref{figfits} we display some synthetic
spectral fits to the observed spectra of four stars of our sample.

\subsection{Ages}

We have gathered together the ages of some of our stars hosting
planets from \citet{Saffe05}.
\citet{Saffe05} studied the correlations between stellar properties 
with age. They measured the chromospheric activity in a sample of 
49 stars with planetary companions and combining with the literature, 
they obtained age estimates for 112 objects. They applied the
calibrations reported in \citet{Donahue93} and \citet{Rocha-Pinto98}
for the chromospheric activity-age relation but also other
methods based on isochrones, lithium abundances, 
metallicities, and kinematics, and they compared them with the 
chromospheric results. 
They concluded that chromospheric activity and isochrone
methods give comparable results, and claimed that isochrone technique
is, in practice, the only tool currently available to derive ages for
the complete sample of planet-host stars, as chromospheric activity 
calibrations of \citet{Donahue93} are limited for ages $<$ 2 Gyr 
and most of the planet-host stars sample are $>$ 2 Gyr age. 
Lithium cannot be used to derived stellar ages greater than 1~Gyr. 
Here, we have adopted the ages derived from isochrones 
when available for objects with age $<$~2~Gyr, otherwise we 
used those derived using the chromospheric activity technique with
\citet{Donahue93} calibrations. The adopted ages are provided in
Table~\ref{tab:par}. We note that the typical uncertainty on the age
determination in field stars is very large, between 1 and 3~Gyr,
especially for dwarf stars.

\begin{figure}
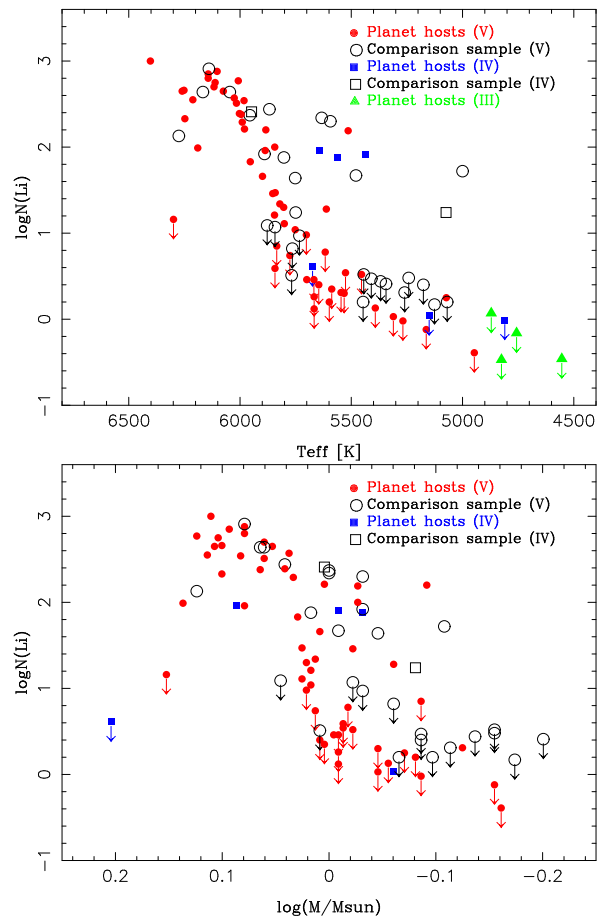

\centering
\includegraphics[width=6.0cm,angle=270,clip]{figure8_go.ps}
\includegraphics[width=6.0cm,angle=270,clip]{limas12o.ps}
\caption{Same as Fig.\ref{figbeteff} but for Li
abundance measurements taken from \citet{Israelian04,Israelian09}.}
\label{figLiteff}
\end{figure}

\section{Discussion}

\subsection{Be and Li abundance versus effective temperature 
and mass \label{secbeliteff}}

In Fig.~\ref{figbeteff} we plot the derived Be abundances as a 
function of effective temperature (top panel) and as a 
function of mass (bottom panel) for the stars in our sample
together with those for the stellar samples reported in
\citet{Garcialopez98,Santos02,Santos04b,Santos04c}.  
We also distinguish among stars with and without planets and 
among different luminosity classes, including
dwarfs (V), subgiants (IV) and giants (III). We note that 
we did not find in the literature any determination of the stellar
mass for all stars in this study, as for instance, no stellar 
mass was available for subgiant and giant planet-host stars with 
$T_{\rm eff}<4900$~K. 

Subgiant and especially giant stars have probably changed their
effective temperatures considerably from their ``initial'' value on
the main sequence and their convective envelopes have also expanded
and reached deeper and hotter regions in the stellar interiors. 
Extra mixing may have already occurred and thus, the dilution
and/or depletion of their light elements do not follow the normal
trend. Therefore, we will take evolved objects into account
separately (see Sect.~\ref{secevol}). 

Considering only dwarfs stars, the addition of the new Be
abundance of 26 planet-host stars in Fig.~\ref{figbeteff} 
shows no significant changes to what \citet{Santos04c} found.
The overall impression is that no clear difference seems to exist
between the two stellar populations, planet hosts and ``single''
stars. 

Be abundances decrease from a maximum near $T_{\rm eff}=6100$~K
towards higher and smaller temperatures, similarly to the behaviour of
Li abundances versus effective temperature (see Fig.~\ref{figLiteff}). 
The steep decrease with increasing temperature resembles the well
known Be gap for F stars \citep[e.g.][]{Boesgaard02}, 
while the decrease of the Be content towards lower temperatures is
smoother, and may show evidence for continuous Be burning
during the main-sequence evolution of these stars 
\citep[see][and references therein]{Santos04c}.

In the top panel of Fig.~\ref{figbeteff} we have also 
posed a set of Yale beryllium-depletion isochrones from 
\citet{Pinsonneault90}. We depict the depletion isochrones 
(case A) for solar metallicity and an age of 1.7 Gyr, 
the standard model and 4 models with different initial
angular momentum \citep[see Table~3-6 of][]{Pinsonneault90} assuming 
an initial $\log N({\rm Be})=1.26$ (intermediate value between solar,
1.10, and meteoritic, 1.42) for all the stars \citep[see][and references therein]{Santos04b}.
As already noticed in \citet{Santos04b,Santos04c}, these models agree
with the observations above roughly 5600~K, but while the observed Be
abundance decreases towards lower temperatures when 
$T_{\rm eff}<5600$~K, these models predict either constant or 
increasing Be abundances.
Several possibilities were also discussed for this
discrepancy between models and Be observations: 
the increasing difficulty to measure Be abundances
at low temperatures; the lack of near-UV line-opacity in the
spectral synthesis; the presence of the planetary companions;
possible accretion of $\approx$0.5$M_{\odot}$ by stars in solar
neighborhood \citep{Murray01}; and possible correlation with oxygen
abundances.  

All of them dismissed as the main possible cause although we will
further discuss the last one in Sect.~\ref{secoxygen}. 
The better explanation is probably that these models do not
predict correctly the behaviour of the Be abundances at the lowest
temperatures. 

Mixing by internal gravity waves may provide an explanation to the
Li and Be depletion in cool dwarf stars 
\citep{Garcialopez91,Montalban94,Montalban96,Charbonnel05}. 
In Fig.~\ref{figbeteff} we also depict one model, that includes 
mixing by gravity waves, from \citet{Montalban00}.
\citet{Santos04b} already pointed out that models 
including mixing by internal waves \citep{Montalban00}, reproduce
the decrease of Be content in the lower temperature regime although
still overestimate the Be abundances with respect to the observed
values. 

 In the bottom panel of Fig.~\ref{figbeteff} we display the
Be abundances as a function of the stellar masses. The Be abundances
show, as expected, a similar trend as compared to that of the top
panel of Fig.~\ref{figbeteff}. However, some stars move with respect
to the other stars in the figure, specially the evolved stars.
As already mention in Sect.~\ref{secbeliteff}, the four giants and 
one subgiant with $T_{\rm eff}<4900$~K do not have available mass 
determinations. 
A priori, one would expect to find the subgiant stars at high
masses, but this picture changes when having stars with different
metallicities. Some subgiant stars with planets stay with relatively
low masses due to their low metal content, whereas other subgiants,
like HD~38529 with a mass of~1.6~$M_\odot$ and [Fe/H]$=0.4$, and with 
an upper-limit Be measurement, have very high masses partially due to 
their high metallicity. 
In the following, we will concentrate on the Be abundances in 
main-sequence stars.

\subsubsection{Main-sequence stars \label{secdwarf}}

In top panel of Fig.~\ref{figbefit} we display Be abundances
as a function of effective temperature only for unevolved stars 
with and without known planetary companion. Here we have 
excluded the stars with Be abundances below 0.4~dex which lie in 
some kind of Be-gap \citep[see][]{Santos04b}.
We perform a linear fit, also overplotted in 
Fig.~\ref{figbefit}, and get a correlation coefficient of
0.74 with standard deviation of 0.15. 
If one discriminates between the planet hosts and the
comparison sample dwarf stars, no significant difference in slope 
is found when fitting all stars in the whole temperature range. 

In Fig.~\ref{figbefit} (top panel) one can see that whereas in 
the range $5700 \lesssim T_{\rm eff}\lesssim 6200$~K, the stars 
with and without planets mostly show similar abundances, at the 
temperature range $5100 \lesssim T_{\rm eff}\lesssim 5500$~K there 
seems to be some indication that planet-host stars may be more Be 
depleted than ``single'' stars. 

\citet{Israelian09}, based in the Li study of unbiased sample of
solar-analogue stars with and without detected planets, find that
planet host have lower Li abundances, being most of them 
upper-limit measurements, than "single" stars, suggesting
that the presence of planets may increase the amount of mixing and
deepen the convective zone to such an extent that the Li can be
burned. 

If this is true, we should see the same effect at lower temperatures
in Be burning regime. This may be reflected in the apparent
overabundance of Be in "single" stars respect to planet host stars 
at 5100--5500~K $T_{\rm eff}$ range, in the sense that most of the Be
measurements in planet hosts are upper-limits.

However, the sample with $T_{\rm eff}$ below 
$5500$~K is still too small. In addition, as shown in Sect.~\ref{secBeLi},
the expected error on the Be abundances in this $T_{\rm eff}$ range
is 0.12-0.17~dex.
Therefore, this calls for new observations to add more Be 
measurements in dwarf stars at these cool temperatures. 

 In Fig.~\ref{figbefit} (bottom panel) one can see that many
points have move to a different relative position according to 
the effective temperature due to the different
metallicity of the stars with and without planets. In particular, 
the stars without planets which on average have lower metallicity
appear to have lower masses. So, the previous picture with the 
points well concentrated in a relatively narrow $T_{\rm eff}$ range,
are now spread in a relatively larger mass range. 
In particular, the main-sequence stars with planets in the 
$T_{\rm eff}$ range 5100--5500~K have a mean metallicity of
[Fe/H]$~\sim0.18$ (with a $\sigma_{\rm Fe}=0.22$) whereas ``single'' stars 
have a mean [Fe/H]$~\sim -0.22$ ($\sigma_{\rm Fe}=0.20$). This
explains why the slight signature in stars with and without planets
seen in the $T_{\rm eff}$ plot gets smudged in the 
$\log(M/M_\odot)$ plot. We note that the star BD--103166, at 
$T_{\rm eff}\sim  5320$~K and with $\log N({\rm Be})<0.5$,
is not display in Figs.~\ref{figbeteff} and~\ref{figbefit}.
We did not find any mass determination for this star, but we may 
estimate its mass to be roughly 0.91~$M_\odot$, and thus having
$\log(M/M_\odot)\sim -0.04$.

 On the other hand, the stellar masses trace the stellar 
positions at the beginning of the stellar life and not at their 
present states, which are established by the stellar temperatures.
As we notice in Sect.~\ref{secfe}, the stellar metallicity may have 
an impact on the Be burning rate but we do not expect that
this explains why most of the Be measurements at the $T_{\rm eff}$ 
range 5100--5500~K in planet hosts are only upper-limits.
Unfortunately, we cannot track the effect of metallicity in the
pre-main sequence and main-sequence evolution of Be abundances 
from theoretical models at different metallicities since
most of the standard models do not predict any Be depletion at 
these cool temperatures irrespective of the metal content of the
star \citep[see e.g.][]{Siess00}.

In Fig.~\ref{figLiteff} we display the Li abundances of the stars in
the sample versus effective temperature  (top panel) and 
versus mass (bottom panel). All the stars in this plot follow the 
general trend: stars with low Be abundances have also their 
Li severely depleted. This general trend seen in the top
panel of Fig.~\ref{figLiteff}, is less clear in the bottom panel due
to the larger spread of the Li abundance measurements in the 
$\log (M/M_\odot)$ axis.
However, there is a small group of both stars with and
without planets at temperatures $T_{\rm eff}\lesssim 5670$~K which
show relatively high values of Li abundance, $\log N({\rm Li})> 1.2$, 
whereas their Be abundances are close to the solar value, $\log
N({\rm Be})\sim 1.1$. Two of these stars are dwarf 
planet-host stars: HD~1237, with a relatively high chromospheric 
activity index, $\log R_{HK}= -4.496$ \citep{Gray06}, 
classified as an active star, and may be a young object, with an age
below 1Gyr; and HD~65216, with a slightly high Li abundance, 
$\log N({\rm Li})= 1.28$, and a low chromospheric index, 
$\log R_{HK}=-4.92$. We note that the 
Sun has $\log R_{HK} \sim -4.75$.
Among these stars there are also four ``single'' dwarf stars: 
three of them, HD~36435, HD~43162, and HD~74576, with also high 
chromospheric activity indexes, $\log R_{HK}= -4.499$,~$-4.480$
and~$-4.402$, respectively. However, the other object, HD~43834, show
a relatively low activity ($\log R_{HK}= -4.940$). 
This star, with a $T_{\rm eff}=5594$~K,  has a ``normal'' Be abundance, 
$\log N({\rm Be})=0.94$~dex and a high Li abundance, $\log N({\rm
Li})=2.30$~dex, for its temperature, according to the general trend.
This makes this star interesting in the context of pollution. 
We refer here the studies by \citet{Santos04b,Santos04b} where 
these four ``single'' stars have been already discussed.

It is also worth remarking one planet-host dwarf star of our 
sample, HD~142, with a temperature $T_{\rm eff} \approx 6400$~K, 
with higher Be content than expected from the typical trend, being
nearly as high as the maximum at $T_{\rm eff} \sim 6100$~K.
This star also has a higher Li abundance than expected
from its effective temperature (see Fig.~\ref{figLiteff}).
We also note that the Be abundance maximum may need to be slightly
shifted towards lower temperatures by one of the planet-host stars
presented in this work, HD~213240, which is located at $T_{\rm
eff}=5982$~K and have $\log N({\rm Be})\sim 1.4$. In other words, one
would have to say that stars with maximum Be abundances have
temperatures in the range $5950 \lesssim T_{\rm eff} \lesssim 
6150$~K. 

\begin{figure}
\centering
\includegraphics[width=6.0cm,angle=270,clip]{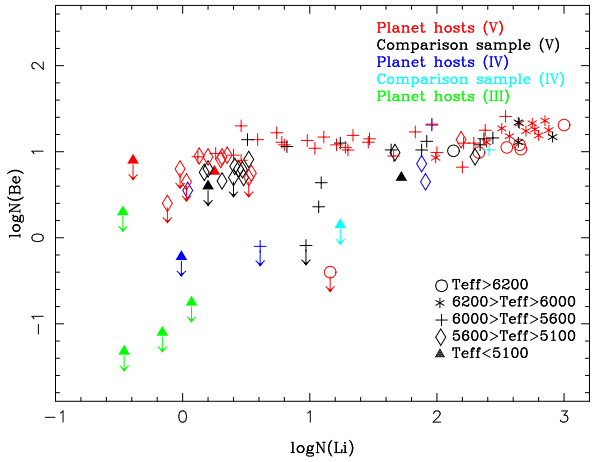}
\caption{Be versus Li abundances of the stars in the sample with
and without planets. Different colors represent different 
luminosity classes. The stars are splitted into different 
temperature ranges associated with different symbols.}
\label{figbeLi}
\end{figure}

\begin{figure}
\centering
\includegraphics[width=6.0cm,angle=270,clip]{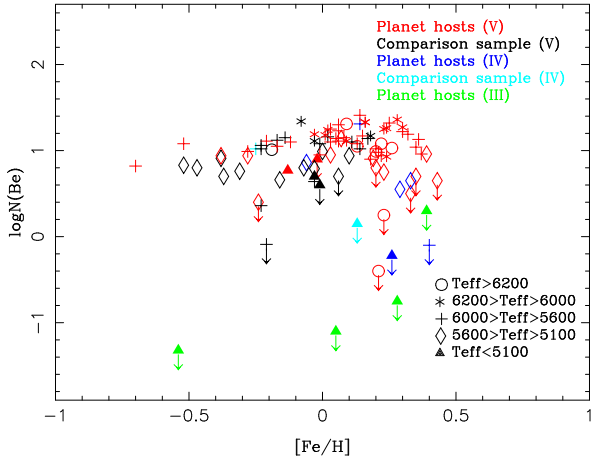}
\caption{Same as Fig.~\ref{figbeLi} but for Be abundance versus 
metallicity.}
\label{figbefe}
\end{figure}

The simplest explanation may be that HD~142, classified as F7V by 
 SIMBAD\footnote{The SIMBAD database can be accessed at http://simbad.u-strasbg.fr/simbad/}, 
 is just a young star which may be supported by its derived and relatively low surface 
 gravity, $\log g = 4.62$.
However, according to the age estimates in \citet{Eggenberger07},
this star has an age of $\approx$2.8-5.93 Gyr which still makes 
lithium content too high. In addition, this star has a 
relatively low chromospheric activity index $\log R_{HK}= -4.853$
\citep{Gray06}, which may indicate that, in fact, the age determined
by \citet{Eggenberger07} may be correct. \citet{Gray06} classified 
HD~142 as a chromospherically inactive star. 
HD~142 lies in the called Be "gap" according to its derived effective
temperature of $T_{\rm eff} \sim 6400$~K, but has nearly keep all 
its Be. In the literature one can find stars with similar or
even higher Be content but typically at $T_{\rm eff} \lesssim 6350$~K,
for instance, in the Hyades cluster with an age of 600~Myr
\citep{Boesgaard02} and in the intermediate-age open cluster 
IC~4651 \citep{Smiljanic10} with an age of 1.7~Gyr, both with a 
metallicity of $\sim0.1$~dex, very similar to the metallicity of
HD~142. 
 We note, however, that the mass of HD~142, 1.29~$M_\odot$,
places this star very close to other main-sequence stars in the 
bottom panel of Figs.~\ref{figbeteff} and~\ref{figbefit}.

New observations of Be in stars with and without planets are also 
needed in this temperature range to study possible pollution effects.
We note that the depletion rates may be different for different
proto-planetary disc masses and composition, and therefore, the
rotational history of the star \citep[e.g.][]{Balachandran95,
Chen01}. Thus we will need to find targets with
similar conditions, i.e. temperature, age, etc. 

\subsubsection{Evolved stars \label{secevol}}

Following the criterion established in Sect.~\ref{secsample}, 
we have four objects clearly classified as giants and 
 eight that could also have evolved off the main sequence.

Five of these subgiants have Li abundance measurements and 
four of them (the planet hosts HD~10697, HD~88133 and HD~117176 and 
the ``single'' star HD~23249) show relatively high values, with
respect to the Li trend described by most of the dwarf stars. 
However, these four subgiant stars have Be abundances consistent 
with the Be trend shown in dwarf stars (see Fig.~\ref{figbeteff}).

HD~10697 and HD~117176, as well as the ``single'' star 
HD~23249 were already discussed in \citet{Santos02,Santos04c}. 
These authors discussed the possibility that pollution by invoking
planet engulfment \citep[see e.g.][]{Israelian01,Israelian03}, 
at least for the two planet hosts, could explain 
the high Li content and the relative ``normal'' Be content
\citep{Siess99}.
However, recent models show that many accretion events of
planetary material could cause Li depletion instead of Li 
enhancement \citep[see e.g.][]{Theado10,Baraffe10}.

The high Li content and also moderate Be in the metal-rich planet
host HD~88133, [Fe/H$]=0.33$, may be also explained by pollution 
effects.  

The four giant stars, HD~47536, HD~59686, HD~219449, and 
HD~27442, are all of them planet hosts and only present upper 
limits for Li and Be. 
 
\subsection{Beryllium versus lithium}

A beryllium versus lithium diagram can provide information on their
different depletion rates in main-sequence stars. 
We have already introduced the relationship between Be and Li
abundance in the previous Sect.~\ref{secbeliteff}, but here,
we will focus on Fig.~\ref{figbeLi}, which depicts the Be abundances
versus Li abundances of the stars in our sample. 
In general, there seems to be almost constant relation between the Be
and Li abundances in dwarf stars, except for very few dwarfs in
the comparison sample. 
However, both stars with and without planets seem to follow 
the same behaviour.

\begin{figure}
\centering
\includegraphics[width=6.0cm,angle=270,clip]{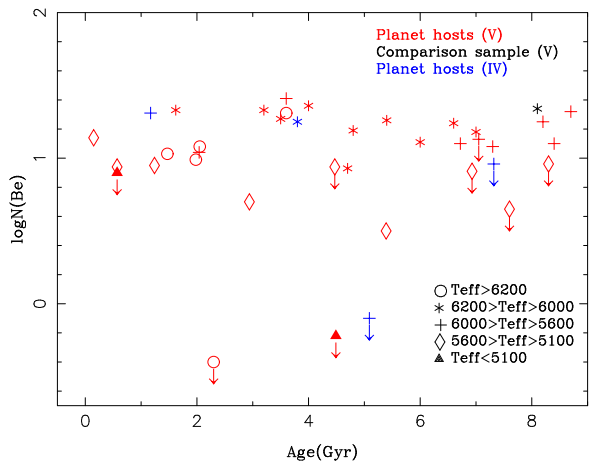}
\caption{
Same as Fig.~\ref{figbeLi} but for Be abundance versus age.}
\label{figbeage}
\end{figure}

In Fig.~\ref{figbeLi} we also split the Be and Li measurements 
in several stellar temperature ranges. Thus, the objects with
$T_{\rm eff} \gtrsim 6000$~K are situated in the upper right
corner, and define the abundance maximum, for both Be and Li, and 
follow a positive correlation.  
The stars with temperatures between 5600 and 6000 K are situated in
the middle of the diagram, in a nearly flat Be abundance through
Li abundances ranging from 0.05 to 2.5~dex. 
The objects with temperatures between 5600 and 5100 K are mostly in
the left upper-middle panel, also follow an almost constant Be
abundance at $\log N({\rm Be})\sim 0.9$ with only upper-limit
Li measurements. 
This temperature range involve all those stars where Be depletion 
has already taken place and therefore Li is severely depleted. 
Thus, this group shows a Be level lower than that at higher 
temperatures. 
Nevertheless, 5 objects in this temperature regime (two of
them are subgiants), both with and without planetary companions, 
are located in the right upper side showing both high levels of Be 
and Li abundances. 
Finally, we separate the objects with temperatures below 5100~K as Be
abundances cannot be determined accurately. These stars can be in turn
divided in three sets: (i) low Be and Li abundances that
correspond to the giant stars; (ii) 
 low Li abundances but
still relatively high Be abundances; and (iii) intermediate abundances
of both Li and Be. The latter contains two objects, one of them
being a subgiant (see Sect.~\ref{secevol}).

Fig.~\ref{figbeLi} confirms what found in \citet{Santos04c}, that Be
and Li burning seem to follow the same trend for the hottest and
coolest objects. Li and Be are depleted in a similar way up to 5600~K,
where Be burning stops while Li burning continues up to $\sim 6000$~K, 
where again both Li and Be are destroyed in a similar way. 
Li burning starts to be severe from $T_{\rm eff}$$\lesssim$5900 K 
while Be keeps its low-burning rate until $T_{\rm eff}$$\sim$5600 K.  
Giant stars show both Li and Be severely depleted with exception of
one object, HD~27442, that still keeps a relatively high upper limit 
to its Be abundance.

In general, Li is more depleted than Be in stars with smaller 
effective temperatures and all stars with and without planetary
companions show a similar behaviour. With the addition of the 26
planet-host stars and one ``single'' star, the apparent ``lack'' 
of planet-host stars in the range of Li abundances 
$\log N({\rm Li}) \sim 1.5-2.5$, noted by \citet{Santos04c}, 
has almost disappeared. 
There seems to be roughly the same number of stars with and without 
planets in this range of Li abundances, but this might also be due 
to the small number of stars without planets, 30, in this 
sample, in comparison with the 70 planet-host stars.
 
\begin{figure}
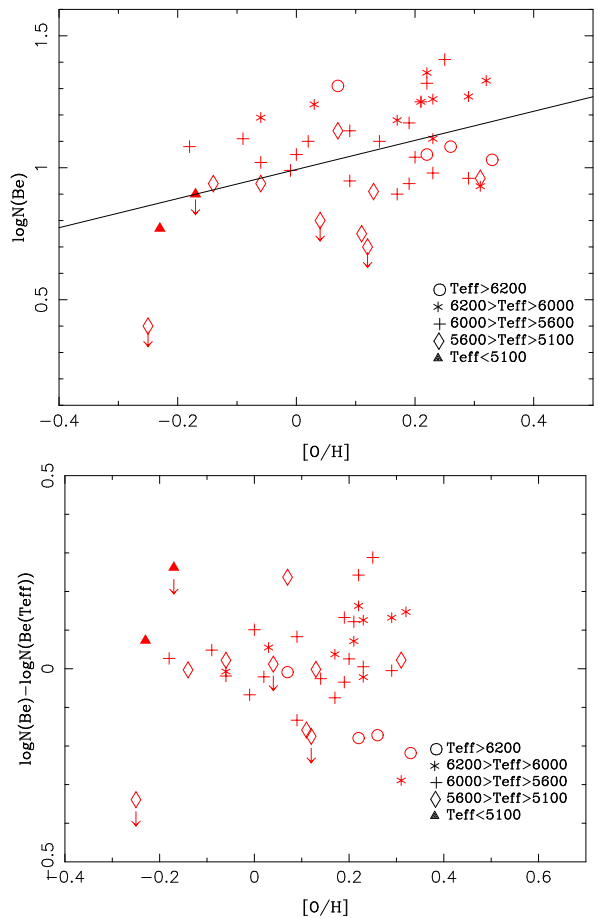

\centering
\includegraphics[width=6.0cm,angle=270,clip]{beox1c6o.ps}
\includegraphics[width=6.0cm,angle=270,clip]{beox2b.ps}
\caption{ {\it Top}: Be versus [O/H] abundances in dwarf 
stars hosting planets in the sample and the best linear fit. 
{\it Bottom}: Be abundance corrected using the linear relation 
Be-$T_{\rm eff}$.
}
\label{figbeo}
\end{figure}

\subsection{Be abundance versus [{\rm Fe}/{\rm H}] 
and age\label{secfe}}

In Fig.~\ref{figbefe} we display the Be abundances versus the metal
content of the stars with and without planets. 
We do not see any clear correlation but just an increasing 
dispersion of the Be content towards higher metallicities. 
The differences in the Be content are mainly due to the differences 
in $T_{\rm eff}$, being lower at lower temperatures, but roughly 
constant for all metallicities. This was already noticed by 
\citet{Santos04c}.

At the highest metallicities, the opacity of the convective zone
changes considerably, producing a higher rate of Li burning and also,
although less significant, Be burning. This might explain why the 
Be abundance in planet-host stars has a maximum at [{\rm Fe}/{\rm H}]~$\sim
0.20$ and then decreases towards higher metallicities. 
On the other hand, Galactic Be abundances are known to increase with
the metallicity \citep[e.g.][]{Rebolo88,Molaro97,Boesgaard99}, which
in fact is seen in Fig.~\ref{figbefe},
although the dispersion in Be abundances and the effective
temperature spread is probably masking this Galactic effect
\citep[see also][and references therein]{Santos04c}.

In Fig.~\ref{figbeage} we display the Be abundances versus ages 
of the planet-host stars including those from the literature.
We have objects in the same temperature regime but with
different ages. There is no correlation, indicating that 
the age, as well as the metallicity, is a secondary effect on the
level of the Be content behind a primary effect driven by the 
effective temperature of the star.
However, any interpretation from this plot must be taken with caution
since the age determinations, specially for main-sequence stars, 
have large error bars and may not be reliable.

\subsection{Be versus [O/H]\label{secoxygen}}

Beryllium is mainly produced in the spallation of CNO nuclei 
 \citep[see e.g.][and references therein]{Tan09}. 
In particular, oxygen gives the largest contribution to the spallation
process responsible for forming Be. Thus, a comparison between O and
Be abundances can provide alternative information on the behaviour of
beryllium. We have collected the oxygen abundances in stars hosting
planets of our sample from \citet{Ecuvillon06}. 
They found average oxygen abundances, [O/H], 0.1-0.2 dex 
higher in stars hosting planets with respect to stars without planets. 
However, although planet accretion was not excluded, 
the possibility that it is the main source of the observed oxygen
enhancement in planet-host stars is 
unlikely. Thus, it seems to be related to the higher metal
content of planet-harbouring stars and both Galactic trends of stars
with and without planet are nearly indistinguishable.

In  (top panel) of Fig.~\ref{figbeo} we display the beryllium 
versus oxygen abundances relative to the Sun only in dwarf 
planet-host stars.
We took the [O/H] values derived from the near-UV OH bands 
in the spectral range 3167--3255~{\AA} \citep{Ecuvillon06}, 
because these lines are located close to the \ion{Be}{ii}~3131~{\AA} 
line and therefore would be affected for the same continuum opacity. 
We note that the stellar parameters used in this work to derive Be
abundances are not exactly the same as those used by
\citet{Ecuvillon06} to derive oxygen abundances, but in general the
differences are very small. We adopted the mean value derived from 
the four OH features as given in \citet{Ecuvillon06}. We 
also note that for some stars in Fig.~\ref{figbeteff} there is no
available O measurement.

The Be content is stratified in different values as the temperature
changes. 
In the temperature range 5100--5600 K Be abundance versus [O/H] 
seems to have a positive slope, just due to one dwarf star located 
at [O/H]~$\sim -0.25$. 
However, all these trends are affected by the strong influence of the
effective temperature on the Be abundances in each star, irrespective
of metallicity and oxygen abundance. 

 In the top panel of Fig.~\ref{figbeo}, similarly to 
Fig.~\ref{figbefit}, we have also displayed the best linear fit, 
showing a correlation of 0.44 with a standard deviation of 0.18.
To study the influence of the effective temperature in this 
Be-O relation, we subtract from each Be measurement the Be content 
given by the Be--$T_{\rm eff}$ linear relation in Fig.~\ref{figbefit}.
We thus obtain the Be abundance "corrected" from the effect of the
effective temperature, $logN({\rm Be})-logN({\rm Be}(T_{\rm eff}))$,
which is shown in the bottom panel of Fig.~\ref{figbeo}.  
The Be content does not seems to depend much on the initial oxygen
abundance of the star, as the trend disappears when removing 
the $T_{\rm eff}$ effect.

\section{Summary and Conclusions}

 We present here new Be abundances of 26 stars harbouring planets
and one star without detected planets, from unpublished 
UVES spectra, added to the previous sample
 of \citet{Garcialopez98,Santos02,Santos04b,Santos04c}. We
have also compiled Li abundances for these objects. The complete
sample contains 100 objects, including dwarfs, subgiants and giants, 
with 70 stars hosting planets. We study the behaviour of Be
abundances of the stars in the sample with effective temperature, Li
abundance, metallicity, age, oxygen abundance and
through different spectral types and luminosity classes.

In general, the work conclude that the beryllium content in stars with and without 
planetary companion is similar and behaves in the same way.
As the burning process is sensitive to the mixing mechanism, the
presence of a planetary companion is not necessary affecting this 
process in a notorious way, at least, at $T_{\rm eff} \gtrsim 5500$~K.

Below $T_{\rm eff} \approx 5500$~K, the stellar sample
is still small but appears to be an increase in the dispersion of
the Be abundances of dwarf stars. In addition, most of the
planet-host Be measurements are not indeed measurements
but upper limits. However, there is no way to check if the
apparent higher scatter of Be content at $T_{\rm eff} \lesssim
5500$~K is due to the different stellar metallicities since most of
the standard theoretical models predict no Be depletion at these
temperatures for any metallicity.

Be depletion depends on the effective temperature more
than on the age and metal content.
The fact that the stars which host planets are richer in
different metal contents than ``single'' stars, but have the
same Be content, supports the idea of primordial origin of these
over-abundances in planet hosts.

\begin{acknowledgements}

M.C.G.O. acknowledges financial support from the European
Commission in the form of a Marie Curie Intra European Fellowship
(PIEF-GA-2008-220679) and the partial support by the Spanish MICINN under
the Consolider-Ingenio 2010 Program grant CSD2006-00070:
First Science with the GTC (http://www.iac.es/consolider-
ingenio-gtc). 
N.C.S. would like to thank the support by the European 
Research Council/European Community under the FP7 through a 
Starting Grant, as well from Funda\c{c}\~ao para a Ci\^encia e a
Tecnologia (FCT), Portugal, through a Ci\^encia\,2007 
contract funded by FCT/MCTES (Portugal) and POPH/FSE (EC), 
and in the form of grant reference PTDC/CTE-AST/098528/2008
from FCT/MCTES.
J.I.G.H. thanks financial support from the Spanish Ministry project
MICINN AYA2008-00695. E.D.M and G.I. would like to thank financial
support from the Spanish Ministry project MICINN AYA 2008-04874.
This research has made use of the SIMBAD database,
operated at CDS, Strasbourg, France. This work has also made use of
the IRAF facility.  

\end{acknowledgements}


\begin{table*}

\caption[]{Stellar parameters and Be abundances measured in this
work. We measured Li abundances for 4 stars. Li abundances without a
label were taken from \citet{Israelian09}.
\label{tab:par}}
\begin{flushleft}
\scriptsize
\begin{tabular}{lccccccclccccc}
\noalign{\smallskip}
\hline
\noalign{\smallskip}
Name     &   $ T_{eff}$  &  log $g$  & $\xi_t$ & [Fe/H]  & Mass & $V_{\rm sini}$  & V & SpT$^{d}$ & LC$^{e}$ & logN(Be) & logN(Li) & [O/H]$^{f}$ & Age$^{g}$ \\
 & [K] & [cm s$^{-2}$]&  km s$^{-1}$ & & M$_{\odot}$ & km s$^{-1}$ & & &  & [dex] & [dex] & {[dex]} & Gyr\\
\noalign{\smallskip}
\hline
\noalign{\smallskip}
HD142$^{a}$  &  6403  &  4.62  &  1.74  &  0.09      & 1.29 &  11.3  &  5.70 &  F7V & V  & 1.31  & 3.00 & 0.07$\pm$0.13 & 3.6  \\
HD1237$^{a}$  &  5514  &  4.50  &  1.09  &  0.07     & 0.94 &  5.5  &  6.59 &  G8.5Vk: & V   & 1.14 & 2.19 & 0.07$\pm$0.10 & 0.15\\
HD4208$^{a}$  &  5599  &  4.44  &  0.78  &  --0.28   & 0.83 & 2.8  & 7.79 &  G7VFe-1H-05 & V  & 0.94 & $<$0.20 & --0.14$\pm$0.08 & 4.47\\
HD23079$^{a}$ &  5980  &  4.48  &  1.12  &  --0.12   & 1.01 & 3.4  & 7.12 &  F9.5V & V & 1.10 & 2.21 & 0.02$\pm$0.09  &  8.4  \\
HD28185$^{a}$  &  5667  &  4.42  &  0.94  &  0.21    & 0.98 & 2.5  & 7.80 &  G6.5IV-V & V  & 0.98 & $<$0.26 & 0.23$\pm$0.10 & 12.2\\
HD30177$^{b}$  &   5588  &  4.29  &  1.08  &  0.39   & 1.01 & 2.96$^{1}$  &  8.41  & G8V & V & 0.96 & $<$0.35$^{2}$ &  0.31$\pm$0.12 & 8.30\\
HD33636$^{b}$*  &   6046  &  4.71  &  1.79  &  --0.08  & 1.16 & 3.08$^{1}$  & 7.06 &  G0VH--03 & V   & 1.34 & 2.64 & 0.02$\pm$0.11 & 8.1\\
HD37124$^{b}$  &   5546  &  4.50  &  0.80  &  --0.38  & 0.75 &  1.22  & 7.68 &  G4IV-V & V & 0.94 & $<$0.31 & --0.06$\pm$0.08 & 0.57\\
HD39091$^{a}$  &   6003  &  4.42  &  1.12  &  0.09   & 1.10 &  3.3 & 5.67 &  G0V & V & 1.11 & 2.39 & 0.23$\pm$0.08 & 6.0 \\
HD47536$^{b}$ &   4554  &  2.28  &  1.82  &  --0.54  & - & 1.93$^{3}$  & 5.26 &  K0III & III  & $<$--1.32& $<$--0.46$^{3}$ & --0.29$\pm$0.12 & --\\
HD50554$^{b}$  &   6026  &  4.41  &  1.11  &  0.01   & 1.09 &  3.88$^{1}$  & 6.84 &  F8V  & V  & 1.18 & 2.57 & 0.17$\pm$0.09 & 7.0 \\
HD59686$^{c}$   &   4871  &  3.15  &  1.85  &  0.28  & - & 0.96$^{3}$  & 5.45 &  K2III & III & $<$--0.75& $<$0.07$^{3}$ & 0.14$\pm$0.16 & --\\
HD65216$^{a}$  &   5612  &  4.44  &  0.78  &  --0.17 & 0.87 & 2.3  & 7.97 & G5V & V & 1.05 & 1.28 &  0.00$\pm$0.08 & --\\
HD70642$^{a}$  &   5668  &  4.40  &  0.82  &  0.18   & 0.98 & 2.8 & 7.18 &  G6VCN+05 & V  & 0.90 & $<$0.46 & 0.17$\pm$0.09 & 10.2\\
HD72659$^{b}$  &   5995  &  4.30  &  1.42  &  0.03   & 1.16 &  2.21$^{1}$ &  7.46 &  G0V & V & 1.25 & 2.38 & 0.21$\pm$0.09 & 8.2\\
HD73256$^{a}$  &   5526  &  4.42  &  1.11  &  0.23   & 0.97 &  3.1  & 8.08 &  G8IV-VFe+05  & V & 0.75 & $<$0.54 & 0.11$\pm$0.11 & 15.9\\
HD74156$^{b}$  &   6112  &  4.34  &  1.38  &  0.16   & 1.27 &  4.32  & 7.61 & G0 & V & 1.33 & 2.75 & 0.32$\pm$0.08 & 3.2 \\
HD88133$^{c}$  &   5438  &  3.94  &  1.16  &  0.33   & 0.98 &  2.17$^{1}$  & 8.06 & G5IV & IV & 0.65& 1.91$^{3}$ & 0.22$\pm$0.09  & 9.56\\
HD99492$^{c}$  &   4810  &  4.21  &  0.72   &  0.26  & - &  1.36$^{1}$  & 7.57 & K2V &  IV  & $<$--0.22& $<$--0.01 &  0.01$\pm$0.08 & 4.49\\
HD106252$^{b}$  &   5899  &  4.34  &  1.08  &  --0.01  & 1.02 & 1.93$^{1}$  & 7.36 & G0V & V & 0.95  & 1.66 &  0.09$\pm$0.07 & 9.2 \\
HD114729$^{a}$  &   5844  &  4.19  &  1.23  &  --0.28  & 0.94 & 2.3  & 6.69 & G0V &  V & 0.99 & 2.00 & --0.01$\pm$0.07 & 11.9 \\
HD117207$^{a}$  &   5667  &  4.32  &  1.01  &  0.22  & 0.98 &  1.8 & 7.26 & G7IV-V & V  & 0.94  & $<$0.12 & 0.19$\pm$0.09 & 16.1 \\
HD117618$^{a}$  &   5990  &  4.41  &  1.13  &  0.03  & 1.08 &  3.2 & 7.17 & G0V & V & 1.10 & 2.29 & 0.14$\pm$0.09 & 6.72\\
HD213240$^{a}$  &   5982  &  4.27  &  1.25  &  0.14  & 1.21 &  4.0 & 6.80 & G0/G1V & V  & 1.41 & 2.54 & 0.25$\pm$0.08 & 3.60\\
HD216435$^{a}$  &   6008  &  4.20  &  1.34  &  0.24  & 1.33 &  5.9 & 6.03 & G0V & V & 1.26 & 2.77 & 0.23$\pm$0.12 & 5.40\\
HD216437$^{b}$  &   5887  &  4.30  &  1.31  &  0.25  & 1.20 &  2.6 & 6.06 & G1VFe+03 & V  & 1.32 & 1.96 &  0.22$\pm$0.10 &  8.7 \\
HD219449$^{c}$  &   4757  &  2.71   &  1.71  &  0.05  & - &  5.10  & 4.21 & K0III & III & $<$--1.1 & $<$--0.16$^{3}$ & --0.10$\pm$0.13 & --\\
\noalign{\smallskip}
\hline
\noalign{\smallskip}
\end{tabular}\\
$^{a}$ Values taken from \citet{Sousa08}.\\
$^{b}$ Values taken from \citet{Santos04a}.\\
$^{c}$ Values taken from \citet{Santos05}.\\
$^{d}$ Values taken from SIMBAD.\\
$^{e}$ Luminosity class assigned by comparing the stellar parameters
with isochrones from \citet{Girardi99} \\
$^{f}$ Values taken from \citet{Ecuvillon06} \\
$^{g}$ Values taken from \citet{Saffe05}\\
$^{1}$ Data taken from \citet{Fischer05}.\\
$^{2}$ Data taken from \citet{Israelian04}.\\
$^{3}$ Data measured in this work.\\
$*$ stars without planet companion.\\
\end{flushleft}
\end{table*}

\begin{table*}
\caption[]{Stellar parameters and Be and Li abundances 
from literature \citep{Garcialopez98,Santos02,Santos04c,Sousa08}: 
Stars hosting planets. \label{tab:parb}}
\begin{flushleft}
\scriptsize
\begin{tabular}{lcccclccccc}
\noalign{\smallskip}
\hline
\noalign{\smallskip}
  Name     &   $ T_{eff}$        &   log $g$   &  [{\rm Fe}/{\rm H}] & Mass$^{a}$ & SpT$^{b}$ & LC$^{c}$  &  logN(Be) & logN(Li) & [O/H]$^{d}$ & Age$^{e}$ \\
 & [K] & [cm s$^{-2}$]& & M$_{\odot}$&  & & [dex] & [dex] & {[dex]} & Gyr\\
\noalign{\smallskip}
\hline
\noalign{\smallskip}
BD--103166 & 5320 & 4.38 & 0.33       & - &  K0V &  V  &  $<$0.5  &  -- & -- & 5.39\\
HD6434  &  5835  &  4.6  &  --0.52    & 0.82 &  G2/G3V & V  &  1.08  &  $<$0.85 & --0.18$\pm$0.1 & 13.3  \\
HD9826  &  6212  &  4.26  &  0.13     & 1.30 &  F8V  & V &  1.05  &  2.55  & 0.22$\pm$0.12 & -- \\
HD10647  &  6143  &  4.48  &  --0.03  & 1.20 &  F9V & V &  1.19  &  2.8 & --0.06$\pm$0.09 & 4.8 \\
HD10697  &  5641  &  4.05  &  0.14    & 1.22 &  G5IV & IV &  1.31  &  1. 96 & -- & 1.17 \\
HD12661  &  5702  &  4.33  &  0.36    & 1.05 &  K0V  & V &  1.13  &  $<$0.98  & -- & 7.05\\
HD13445  &  5163  &  4.52  &  --0.24  & 0.70 &  K1V  & V &  $<$0.4  &  $<$--0.12 & --0.25$\pm$0.12 & -- \\
HD16141  &  5801  &  4.22  &  0.15    & 1.06 &  G5IV & V  &  1.17  &  1.11  & 0.19$\pm$0.10 & 11.2\\
HD17051  &  6252  &  4.61  &  0.26    & 1.26 &  F9VFe+03 & V &  1.03  &  2.66 & 0.33$\pm$0.11 & 1.47 \\
HD19994  &  6190  &  4.19  &  0.24    & 1.37 &  F8V  & V &  0.93  &  1.99 & 0.31$\pm$0.12  & 4.7  \\
HD22049  &  5073  &  4.43  &  --0.13  & 0.85 &  K2V  & V &  0.77  &  $<$0.25 & --0.23$\pm$0.10 & --  \\
HD27442  &  4825  &  3.55  &  0.39    & - &   K2III  & III &  $<$0.3  &  $<$--0.47 & 0.07$\pm$0.14 & -- \\
HD38529  &  5674  &  3.94  &  0.4     & 1.60 &   G4V  & IV &  $<$--0.1  &  $<$0.61 & --  &  5.09\\
HD46375  &  5268  &  4.41  &  0.2     & 0.82 &   K1IV & V  &  $<$0.8  &  $<$--0.02  & 0.04$\pm$0.10 & 16.4\\
HD52265  &  6103  &  4.28  &  0.23    & 1.20 &  G0III-IV & V &  1.25  &  2.88  & 0.21$\pm$0.10 & 3.8 \\
HD69830  &  5410  &  4.38  &  --0.03  & 0.82 &  K0V  & V &  0.79  &  $<$0.47  \\
HD75289  &  6143   &  4.42  &  0.28   & 1.24 &  F9VFe+03 & V &  1.36  &  2.85  & 0.22$\pm$0.11 & 4.0 \\
HD75732A$^{f}$  &  5150  &  4.15  &  0.29  & 0.87 & G8V & IV &  0.55  &  $<$0.04  &     --         & --\\
HD82943  &  6016  &  4.46  &  0.3     & 1.15 &   F9VFe+05 & V &  1.27  &  2.51  & 0.29$\pm$0.11 & 3.5 \\
HD83443  &  5454  &  4.33  &  0.35    & 0.95 &  K0V & V &  $<$0.7  &  $<$0.52  & 0.12$\pm$0.13 & 2.94\\
HD92788  &  5821  &  4.45  &  0.32    & 1.03 &  G5 & V &  1.19  &  1.34  &         --     & 9.6 \\
HD95128  &  5954  &  4.44  &  0.06    & 1.07 &  G1V & V &  1.23  &  1.83 &        --       & --\\
HD108147  &  6248  &  4.49  &  0.2    & 1.26 &  F8VH+04 & V  &  0.99  &  2.33 &  --              & 1.98 \\
HD114762  &  5884  &  4.22  &  --0.7  & 0.81 &  F9V & V &  0.82  &  2.2  &          --       & 11.8\\
HD117176  &  5560  &  4.07  &  --0.06 & 0.93 &  G5V  & IV &  0.86  &  1.88  &       --         & -- \\
HD120136  &  6339  &  4.19  &  0.23   & 1.33 &   F6IV & V    &  $<$0.25  &  --  & --  &  -- \\
HD121504  &  6075  &  4.64  &  0.16   & 1.13 &   G2V  & V &  1.33  &   2.65  &    --           & 1.62\\
HD130322  &  5392  &  4.48  &  0.03   & 0.88 &   K0III  & V &  0.95  &  $<$0.13 & --  & 1.24\\
HD134987  &  5776  &  4.36  &  0.3    & 1.03 &   G5V  & V &  1.22  &  $<$0.74  &       --        & 11.1\\
HD143761  &  5853  &  4.41  &  --0.21 & 0.95 &   G0Va & V &  1.11  &  1.46  & --0.09$\pm$0.08 & -- \\
HD145675  &  5311  &  4.42  &  0.43   & 0.90 &   K0V  & V &  $<$0.65  &  $<$0.03  &    --           & 7.6 \\
HD168443  &  5617  &  4.22  &  0.06   & 0.96 &   G6V  & V &  1.11  &  $<$0.78  &      --          & 10.6 \\
HD169830  &  6299  &  4.1  &  0.21    & 1.42 &   F7V  & V &  $<$--0.4  &  $<$1.16 & 0.22$\pm$0.12 &  2.3 \\
HD179949  &  6260  &  4.43  &  0.22   & 1.28 &  F8.5V  & V &  1.08  &  2.65 &   0.26$\pm$0.11 &  2.05\\
HD186427$^{f}$  &  5700  &  4.35  &  0.06  & 0.99 &  G3V  & V &  1.3  &  0.46 &       --         & -- \\
HD187123  &  5845  &  4.42  &  0.13   & 1.04 &  G5  & V &  1.08  &  1.21  &       --       & 7.3\\
HD192263  &  4947  &  4.51  &  --0.02 & 0.69 &  K2V & V &  $<$0.9  &  $<$--0.39 & --0.17$\pm$ 0.09 & 0.57  \\
HD195019  &  5842  &  4.32  &  0.08   & 1.06 &   G3IV-V & V &  1.15  &  1.47 &         --      & 10.6\\
HD202206  &  5752  &  4.5  &  0.35    & 1.04 &   G6V  & V &  1.04  &  1.04  & 0.20$\pm$0.09  & 2.04\\
HD209458  &  6117  &  4.48  &  0.02   & 1.15 &  G0V  & V &  1.24  &  2.7  & 0.03$\pm$0.07 & 6.6 \\
HD210277  &  5532  &  4.29  &  0.19   & 0.90 &  G0V  & V &  0.91  &  $<$0.3  & 0.13$\pm$0.12 & 6.93\\
HD217014  &  5804  &  4.42  &  0.2    & 1.05 &   G5V  & V &  1.02  &  1.3  & --0.06$\pm$0.11 & --\\
HD217107  &  5646  &  4.31  &  0.37   & 1.02 &  G8IV & V  &  0.96  &  $<$0.4  & 0.29$\pm$0.13 & 7.32\\
HD222582  &  5843  &  4.45  &  0.05   & 0.97 & G5 & V &  1.14  &  $<$0.59  & 0.09$\pm$0.08& 11.1\\
\noalign{\smallskip}
\hline
\noalign{\smallskip}
\end{tabular}\\
$^{a}$ Values taken from \citet{Sousa08} and \citet{Santos04a}\\
$^{b}$ Values taken from SIMBAD.\\
$^{c}$ Luminosity class assigned by comparing the stellar parameters
with isochrones from \citet{Girardi99} \\
$^{d}$ Values taken from \citet{Ecuvillon06} \\
$^{e}$ Values taken from \citet{Saffe05}\\
$^{f}$ Values taken from \citet{Garcialopez98}\\
\end{flushleft}
\end{table*}

\begin{table*}
\caption[]{Stellar parameters and Be and Li abundances from
 literature \citep[see][]{Garcialopez98,Santos02,Santos04c,Sousa08}:
"Single Stars" \label{tab:parc}} 
\begin{flushleft}
\scriptsize
\begin{tabular}{lcccclccc}
\noalign{\smallskip}
\hline
\noalign{\smallskip}
 Name & $ T_{\rm eff}$ & $\log g$ & [Fe/H] & Mass$^{a}$ &  SpT$^{b}$ & LC$^{c}$ & $\log N$(Be) & $\log N$(Li) \\
 & [K] & [cm s$^{-2}$]&  & M$_{\odot}$ & &  & [dex] & [dex] \\
\noalign{\smallskip}
\hline
\noalign{\smallskip}
HD870  &  5447  &  4.57  &  --0.07   & 0.86 &  K0V  & V &  0.8  & $<$0.2  \\
HD1461  &  5768  &  4.37  &  0.17    & 1.02 &  G0V  & V &  1.14  &  $<$0.51  \\
HD1581  &  5956  &  4.39  &  --0.14  & 1.00 &  F9.5V  & V &  1.15  &  2.37  \\
HD3823  &  5948  &  4.06  &  --0.25  & 1.01 &  G0VFe-09H-04 & IV &  1.02  &  2.41  \\
HD4391  &  5878  &  4.74  &  --0.03  & 1.11 &  G5VFe-08 & V &  0.64  &  $<$1.09  \\
HD7570  &  6140  &  4.39  &  0.18    & 1.20 &  F9VFe+04 & V &  1.17  &  2.91  \\
HD10700  &  5344  &  4.57  &  --0.52 & 0.63 &  G8V & V &  0.83  &  $<$0.41  \\
HD14412  &  5368  &  4.55  &  --0.47 & 0.73 &  G8V & V &  0.8  &  $<$0.44  \\
HD20010  &  6275  &  4.4  &  --0.19  & 1.33 &   F6V & V &  1.01  &  2.13  \\
HD20766  &  5733  &  4.55  &  --0.21 & 0.93 &  G2V & V &  $<$--0.09  &  $<$0.97  \\
HD20794  &  5444  &  4.47  &  --0.38 & 0.70 &  G8V & V &  0.91  &  $<$0.52  \\
HD20807  &  5843  &  4.47  &  --0.23 & 0.95 &  G0V & V &  0.36  &  $<$1.07  \\
HD23249  &  5074  &  3.77  &  0.13   & 0.83 &  K0IV & IV &  $<$0.15  &  1.24  \\
HD23484  &  5176  &  4.41  &  0.06   & 0.82 &  K2Vk: & V & $<$0.70  &  $<$0.4  \\
HD26965A  &  5126  &  4.51  &  --0.31 & 0.67 & K1V  & V &  0.76  &  $<$0.17  \\
HD30495  &  5868  &  4.55  &  0.02   & 1.10 &  G1.5VH-05  & V &  1.16  &  2.44  \\
HD36435  &  5479  &  4.61  &  0.00   & 0.98 &  G9V & V &  0.99  &  1.67  \\
HD38858  &  5752  &  4.53  &  --0.23 & 0.90 &  G4V & V &  1.02  &  1.64  \\
HD43162  &  5633  &  4.48  &  --0.01 & 1.00 &  G6.5V  & V &  1.08  &  2.34  \\
HD43834  &  5594  &  4.41  &  0.1    & 0.93 &  G7V  & V &  0.94  &  2.3  \\
HD72673  &  5242  &  4.5  &  --0.37  & 0.70 &  G9V & V &  0.7  &  $<$0.48  \\
HD74576  &  5000  &  4.55  &  --0.03 & 0.78 &  K2.5Vk: & V &  0.7  &  1.72  \\
HD76151  &  5803  &  4.5  &  0.14    & 1.04 &  G2V & V &  1.02  &  1.88  \\
HD84117  &  6167  &  4.35  &  --0.03 & 1.15 &  F8V & V &  1.11  &  2.64  \\
HD186408$^{d}$  &  5750  &  4.2  &  0.11  & - &  G1.5Vb & V &  1.1  &  1.24  \\
HD189567  &  5765  &  4.52  &  --0.23  & 0.87 &   G2V & V &  1.06  &  $<$0.82  \\
HD192310  &  5069  &  4.38  &  --0.01  & 0.80 &  K2+v & V &  $<$0.6  &  $<$0.2  \\
HD211415  &  5890  &  4.51  &  --0.17  & 0.93 &  G0V & V &  1.12  &  1.92  \\
HD222335  &  5260  &  4.45  &  --0.16  & 0.77 &  G9.5V & V &  0.66  &  $<$0.31  \\
\noalign{\smallskip}
\hline
\noalign{\smallskip}
\end{tabular}\\
$^{a}$ Values taken from \citet{Sousa08} and \citet{Santos04a}\\
$^{b}$ Values taken from SIMBAD.\\
$^{c}$ Luminosity class assigned by comparing the stellar parameters
with isochrones from \citet{Girardi99} \\
$^{d}$ Values taken from \citet{Garcialopez98}\\
\end{flushleft}
\end{table*}


\end{document}